\documentclass[12pt]{article}
\usepackage{amsmath}

\begin{document}
\title{Bianchi type-VI model with cosmic strings in the presence of a
magnetic field}

\author{Bijan Saha
\thanks{E-mail:~~~bijan@jinr.ru;
URL: http://bijansaha.narod.ru/}\\
{\small \it Laboratory of Information Technologies}\\
{\small \it Joint Institute for Nuclear Research, Dubna}\\
{\small \it 141980 Dubna, Moscow region, Russia}
\and
Mihai Visinescu
\thanks{E-mail:~~~mvisin@theor1.theory.nipne.ro;
URL: http://www.theory.nipne.ro/\~\,\!mvisin/}\\
{\small \it Department of Theoretical Physics}\\
{\small \it National Institute for Physics and Nuclear Engineering}\\
{\small \it Magurele, P. O. Box MG-6, RO-077125 Bucharest, Romania}
}
\date{}

\maketitle

\begin{abstract}

A Bianchi type-VI  cosmological model in the presence of a magnetic
flux together with a cloud of cosmic strings is considered.
In general, the presence of a magnetic field  imposes severe 
restrictions regarding the consistency of the field equations. These 
difficulties could be overtaken working either in a Bianchi 
type-VI$_0$ spacetime
or assuming a particular coordinate-dependence of the magnetic field.
Using a few plausible assumptions regarding the parametrization of 
the cosmic strings, some exact analytical solutions 
are presented. Their asymptotic behavior for large time is exhibited.

~

Pacs: 95.30.Sf; 98.80.Jk; 04.20.Ha

\end{abstract}

\section{Introduction}

Since the observation of small anisotropies in the microwave background
radiation (CMB) \cite{CMB} and large scale structures \cite{SDSS} it
became clear that a pure Friedmann-Lemaitre-Robertson-Walker (FLRW)
cosmology could not explain all the properties of our Universe. It is
therefore natural to consider anisotropic cosmological models that
allow FLRW Universes as special cases.

It is usually assumed that at the very early stages of the evolution of
the Universe, during the phase transitions, the symmetry of the
Universe was broken spontaneously \cite{Kibble,Vil}. Topological
defects such as strings, domain walls, monopoles has received
considerable attention in cosmology since they could play an important
role in the formation of large structure of the Universe.

String cosmological models have been used in
attempts to describe the early Universe and to investigate
anisotropic dark energy component including a coupling between dark
energy and a perfect fluid (dark matter) \cite{KM}. Cosmic
strings are one dimensional topological defects associated with
spontaneous symmetry breaking in gauge theories. Their presence in
the early Universe can be justified in the frame of grand unified
theories (GUT).

A large number of astrophysical observations proves the existence of
magnetic fields in galaxies. Galactic magnetic fields which we observe
today could be relics of a coherent magnetic field existing in the
early Universe, before galaxy formation.
Any theoretical study of cosmological models which contain a magnetic
field must take into account that the corresponding Universes are
necessarily anisotropic. Among the anisotropic spacetimes, Bianchi
type-VI (BVI) space seems to be one of the most convenient for testing
different cosmological models.

The object of this paper is to investigate a BVI string
cosmological model in the presence of a magnetic field due to an
electric current together with the strings. Our paper is organized as
follows: In Section 2 we derive the field equations of BVI cosmic
string model in the presence of a magnetic field. Section 3 deals
with the exact solutions obtained using some simple plausible
assumptions and describe their asymptotic behavior. The last Section
contains conclusions. In the Appendix A, the geometrical properties
and the shear tensor for BVI model are briefly described.

\section{Model and field equations}

The gravitational field in our case is given by a BVI metric:
\begin{equation}
ds^2 = dt^2 - a_1^{2} e^{-2mz}\,dx^{2} - a_2^{2} e^{2nz}\,dy^{2} -
a_3^{2}\,dz^2\,, \label{bvi}
\end{equation}
with $a_1,\,a_2,\,a_3$ being the functions of time only. Here
$m,\,n$ are some arbitrary constants and the velocity of light is
taken to be unity. It should be emphasized that the BVI metric
models a Universe that is anisotropic and space-dependent. The
geometrical properties of the BVI spacetime are sketch in Appendix
A, including the relationship with other Bianchi-type Universes. The
Einstein field equations for BVI metric \eqref{bvi} are written in
the form:
\begin{subequations}
\label{ein}
\begin{eqnarray}
\frac{\ddot a_2}{a_2} +\frac{\ddot a_3}{a_3} +\frac{\dot
a_2}{a_2}\frac{\dot
a_3}{a_3} - \frac{n^2}{a_3^2} &=& \kappa T_{1}^{1}\,, \label{11}\\
\frac{\ddot a_3}{a_3} +\frac{\ddot a_1}{a_1} +\frac{\dot
a_3}{a_3}\frac{\dot
a_1}{a_1} - \frac{m^2}{a_3^2} &=& \kappa T_{2}^{2}\,, \label{22} \\
\frac{\ddot a_1}{a_1} +\frac{\ddot a_2}{a_2} +\frac{\dot
a_1}{a_1}\frac{\dot
a_2}{a_2} + \frac{m n}{a_3^2} &=& \kappa T_{3}^{3}\,, \label{33}\\
\frac{\dot a_1}{a_1}\frac{\dot a_2}{a_2} +\frac{\dot a_2}{a_2}
\frac{\dot a_3}{a_3} + \frac{\dot a_3}{a_3}\frac{\dot a_1}{a_1} -
\frac{m^2 - m n + n^2}{a_3^2} &=&
\kappa T_{0}^{0}\,, \label{00}\\
m \frac{\dot a_1}{a_1} - n \frac{\dot a_2}{a_2} - (m - n) \frac{\dot
a_3}{a_3} &=& \kappa T_{3}^{0}\,. \label{03}
\end{eqnarray}
\end{subequations}
Here overdots denote differentiation with respect to time ($t$).
The energy-momentum tensor for a system of cosmic strings and magnetic
field is chosen to be
\begin{equation}
T_{\mu}^{\nu} =  \rho u_\mu u^\nu - \lambda x_\mu x^\nu +
E_\mu^\nu\,, \label{imperfl}
\end{equation}
where $\rho$ is the rest energy density of strings with massive
particles attached to them and can be expressed as $\rho = \rho_{p}
+ \lambda$, where $\rho_{p}$ is the rest energy density of the
particles attached to the strings and $\lambda$ is the tension
density of the system of strings \cite{letelier,pradhan,tade} which
may be positive or negative. Here $u_i$ is the four velocity and
$x_i$ is the direction of the string, obeying the relations
\begin{equation}
u_iu^i = -x_ix^i = 1, \quad u_i x^i = 0\,. \label{velocity}
\end{equation}
For the electromagnetic field $E_{\mu\nu}$ we adopt the form given
by Lichnerowich \cite{lich}
\begin{equation}
E_\mu^\nu = {\bar \mu} \Bigl[ |h|^2 \Bigl(u_\mu u^\nu - \frac{1}{2}
\delta_\mu^\nu\Bigr) - h_\mu h^\nu \Bigr]\,. \label{lichn}
\end{equation}
Here $\bar \mu$ is the magnetic permeability and $h_\mu$ is the
magnetic flux vector defined by
\begin{equation}
h_\mu = \frac{1}{\bar \mu}  \ast  F_{\nu \mu} u^\nu, \label{magflux}
\end{equation}
where $\ast F_{\mu\nu}$ is the dual of the electromagnetic field
tensor $F_{\mu\nu}$.

In what follows the comoving
coordinates are taken to be $u^0 = 1,\, u^1 = u^2 = u^3 = 0$. We
choose the incident magnetic field to be in the direction of
$z$-axis so that the magnetic flux vector has only one nontrivial
component, namely $h_3 \ne 0.$ In view of the aforementioned
assumption from \eqref{magflux} one obtains $F_{23} = F_{31} = 0.$
We also assume that the conductivity of the magnetic fluid is infinite 
which leads to $F_{01} = F_{02} = F_{03} = 0$. Therefore there is only 
one non-vanishing component of $F_{\mu \nu}$, namely $F_{12}$. Then
from the first set of Maxwell equation
\begin{equation}
F_{\mu\nu;\beta} + F_{\nu\beta;\mu} + F_{\beta \mu; \nu} = 0\,,
\label{maxe}
\end{equation}
where the semicolon stands for covariant derivative, one finds
\begin{equation}
F_{12} = {\cal I}, \quad {\cal I} = {\rm const.} \label{f23}
\end{equation}
Then from \eqref{magflux} we get
\begin{equation}
h_3 = \frac{a_3 {\cal I}}{{\bar \mu} a_1 a_2} \exp{[(m-n)z]}\,.
\label{h1}
\end{equation}
Finally, for $E_\mu^\nu$ one finds the following nontrivial
components
\begin{equation}
E_0^0 = -E_1^1 = - E_2^2 =  E_3^3 = E = \frac{{\cal I}^2} {2 {\bar
\mu} a_1^2 a_2^2} \exp{[2(m-n)z]}\,. \label{E}
\end{equation}
Using comoving coordinates we have the following components of
energy momentum tensor \cite{ass}:
\begin{equation}\label{total}
T_{0}^{0} - \rho = - T_{1}^{1} = - T_{2}^{2} =  T_{3}^{3} - \lambda
= \frac{{\cal I}^2} {2 {\bar \mu} a_1^2 a_2^2} \exp{[2(m-n)z]}\,.
\end{equation}
Taking into account that $T_3^0 = 0$ from \eqref{03} one immediately
finds
\begin{equation}
\Bigl(\frac{a_1}{a_3}\Bigr)^m = {\cal N}
\Bigl(\frac{a_2}{a_3}\Bigr)^n, \quad {\cal N} = {\rm const.}
\label{abcrel}
\end{equation}
Let us now introduce a new function
\begin{equation}
v = a_1 a_2 a_3. \label{defv}
\end{equation}
Then from \eqref{abcrel} one finds
\begin{eqnarray}
a_1 &=& \Bigl({\cal N} v^n a_3^{m-2n}\Bigr)^{1/(m+n)}\,, \label{a1}\\
a_2 &=& \Bigl( v^m a_3^{n-2m}/{\cal N}\Bigr)^{1/(m+n)}\,. \label{a2}
\end{eqnarray}

Summation of \eqref{11}, \eqref{22}, \eqref{33} and three times
\eqref{00} gives

\begin{equation}
\frac{\ddot v}{v} = 2 \frac{m^2 - mn + n^2}{a_3^2} +
\frac{\kappa}{2}\Bigl(3 \rho + \lambda + \frac{{\cal I}^2 a_3^2} {
{\bar \mu} v^2 } \exp{[2(m-n)z]}\Bigr)\,. \label{detv}
\end{equation}
 
Let us note that from the energy-momentum conservation law one finds
\begin{equation}
\dot \rho + \frac{\dot v}{v} \rho - \frac{\dot a_3}{a_3} \lambda =
0\,. \label{emc}
\end{equation}

Taking into account the $z$-dependence of the energy-momentum tensor 
\eqref{total} the r. h. s. of eqs. \eqref{ein} have also a 
$z$-dependence, while the metric functions $a_i$ depend on time only. 
Therefore, in general, eqs. \eqref{ein} are inconsistent for a BVI
model in the presence of a magnetic field. 
There are two possibilities to restore the consistency of eqs. 
\eqref{ein}:
\begin{enumerate}
\item{to limit ourselves to Bianchi type-VI$_0$ (BVI$_0$), namely, 
to consider the case $m=n$;}
\item{to assume a special $z$-dependence of the magnetic permeability 
$\bar\mu$ in order to compensate for the $z$-dependence of the magnetic 
flux component $h_3$, eq. \eqref{h1}. A similar assumption was used by 
Bali \cite{Bali} in a different context}
\end{enumerate}

In what follows we shall analyze these two possibilities in turn.

\section{Solutions of field equations}
\subsection{BVI$_0$ model}

Assuming $m=n$ we have
\begin{equation} \label{a1a2}
a_1 = {\cal N} a_2, \quad {\cal N} = {\rm const.}
\end{equation}
which permit us to express $a_1$ and $a_2$ in terms of $a_3$ and $v$:
\begin{eqnarray}
a_1 &=& {\cal N}^{1/2} v^{1/2} a_3^{-1/2}\,, \label{a10}\\
a_2 &=& {\cal N}^{-1/2} v^{1/2} a_3^{-1/2}\,. \label{a20}
\end{eqnarray}

Consequently Einstein's equations \eqref{ein} are reduced to the 
following set of independent equations:
\begin{subequations}
\label{einVind}
\begin{eqnarray}
\frac{\ddot a_3}{a_3} +\frac{\ddot a_1}{a_1} +\frac{\dot
a_3}{a_3}\frac{\dot
a_1}{a_1} - \frac{m^2}{a_3^2} &=& -\frac{{\cal K}}{a_1^4}\,, 
\label{22Vind} \\
2\frac{\ddot a_1}{a_1}  + \frac{\dot a_1^2}{a_1^2}
+ \frac{m^2}{a_3^2} &=& \frac{{\cal K}}{a_1^4}
+\lambda\,, \label{33Vind}\\
\frac{\dot a_1^2}{a_1^2} + 2\frac{\dot a_1}{a_1}
\frac{\dot a_3}{a_3} -\frac{m^2}{a_3^2} &=&
\frac{{\cal K}}{a_1^4} +\rho, \label{00Vind}\,,
\end{eqnarray}
\end{subequations}
where we introduced the notation
\begin{equation}
{\cal K} = \frac{\kappa {\cal I}^2 {\cal N}^2} {2 {\bar \mu}}\,.
\end{equation}

Therefore there are three equations \eqref{einVind} for four unknown 
functions $a_1$, $a_3$, $\rho$, $\lambda$.
It is customary to assume a relation between $\rho$ and $\lambda$ in
accordance with the equations of state for strings. The simplest one is
a proportionality relation \cite{letelier}:
\begin{equation}\label{rhoalphalambda}
\rho = \alpha \lambda \,.
\end{equation}
The  most usual choices of the constant $\alpha$ are
\cite{ass,SV,SRV,IJTP}
\begin{equation}\label{alpha}
\alpha =\left \{
\begin{array}{ll}
1 & \quad {\rm geometric\,\,\,string}\\
1 + \omega  & \quad \omega \ge 0, \quad p \,\,{\rm
string\,\,\,or\,\,\,
Takabayasi\,\,\,string}\\
-1  & \quad {\rm Reddy\,\,\,string}\,.
\end{array}
\right.
\end{equation}

Using relation \eqref{rhoalphalambda} between $\rho$ and $\lambda$
we get that
\begin{equation}\label{rhova3}
\rho v a_3^{-\frac{1}{\alpha}} = C\,,
\end{equation}
with $C$ an arbitrary constant.

Now the system of differential equations is determined and we can 
proceed to solve it. However this system of nonlinear differential 
equations is quite intricate and we should resort to numerical 
simulations which will be reported elsewhere \cite{SRVprep}.

Here we limit ourselves to investigate the asymptotic behavior of the 
solutions for large $t$. For example, we can investigate the possibility 
to reach an isotropic regime, i.e. all functions $a_i$ to have a 
similar behavior for $t\longrightarrow \infty$
\begin{equation}\label{asymp}
a_1 \sim a_2 \sim a_3 \sim v^{\frac{1}{3}}\,,
\end{equation}
and consequently for the density of strings
\begin{equation}\label{rhoasymp}
\rho \sim v^{\frac{1}{3\alpha}-1}\,.
\end{equation}

As it can be observed from \eqref{sh11} - \eqref{sh33} in the case  of 
isotropic spacetime the components of the shear tensor $\sigma_i^i$  
vanish.

To explore this possibility of isotropization, we shall investigate the 
equation of evolution of $v$, \eqref{detv} for $m=n$:
\begin{equation}
\frac{\ddot v}{v} = 2 \frac{m^2}{a_3^2} +
\frac{\kappa}{2}\Bigl(3 \rho + \lambda + \frac{{\cal I}^2 a_3^2} {
{\bar \mu} v^2 } \Bigr)\,. \label{vm=n}
\end{equation}

Assuming the asymptotic relation between $a_i$ and $v$ \eqref{asymp} we 
get from \eqref{vm=n} the following differential equation valid in the 
asymptotic regime $t\longrightarrow \infty$:
\begin{equation}\label{detvasym}
{\ddot v} = C_1 v^{\frac{1}{3}} + C_2 v^{\frac{1}{3\alpha}} +
C_3 v^{-\frac{1}{3}}\,,
\end{equation}
where $C_1=2m^2, C_2 = \frac{\kappa (3 \alpha +1)}{2\alpha},
C_3= \frac{\kappa {\cal I}^2}{2 {\bar \mu}}$ are some constants. 
This equation allows the following first integral
\begin{equation}
\int \frac{dv}{\sqrt{C_4 v^{\frac{4}{3}} + C_5 v^{1+\frac{1}{3\alpha}}
+C_6 v^{\frac{2}{3}} + C_7}} = t + t_0\,, \label{quad1asym}
\end{equation}
where $C_4 = 3 C_1/2$, \quad $C_5 = 6 \alpha C_2/(3\alpha + 1)$ and
$C_6 = 3 C_3$. Here $t_0$ and $C_7$ are constants of integrations.

We observe that for
\begin{equation}
\frac{1}{3\alpha} \leq \frac{1}{3}\,,
\end{equation}
i.e, $\alpha \geq 1$ or $\alpha < 0$ , the term with $v^{\frac{4}{3}}$
is dominant in the integration \eqref{quad1asym} and finally we get
\begin{equation}\label{vasym}
v \sim t^{3}\,,
\end{equation}
and consequently
\begin{equation}\label{a3asym}
a_3 \sim t\,,
\end{equation}
and
\begin{equation}\label{rhoasym}
\rho \sim t^{\frac{1}{\alpha}-3}\,.
\end{equation}
On the other hand for
\begin{equation}
\frac{1}{3\alpha} > \frac{1}{3}\,,
\end{equation}
i.e. $\alpha \in (0,1)$, the term with $v^{1+\frac{1}{3\alpha}}$ is
dominant in the integration \eqref{quad1asym} and we obtain
\begin{equation}\label{vasymp}
v \sim t^{\frac{6\alpha}{3\alpha-1}}\,.
\end{equation}
For $\alpha \in (\frac{1}{3},1)$ we have a power growing in time for
$v$ and in the limiting case $\alpha = \frac{1}{3}$ we get an
exponential behavior in time.
Finally, we note that for $\alpha \in (0,\frac{1}{3})$ there are no
solutions in this model presenting an expansion of the Universe for
large $t$.
\subsection{BVI model with a specific magnetic permeability}

As a second possibility to assure the compatibility of eqs. \eqref{ein} 
we assume a special $z$-dependence of the magnetic permeability 
\cite{Bali}:
\begin{equation}\label{mu}
{\bar \mu} = {\bar \mu_0}\exp{[2(m-n)z]}\,,
\end{equation}
with ${\bar \mu_0}$ a constant. Let us note that for $z=0$ the 
exponential factor is $1$, but for $z\longrightarrow \pm \infty$ 
this factor vanishes or diverges, depending of the sign of the 
difference $m-n$. This unusual behavior of the magnetic permeability 
is accepted here as a working hypothesis.

In this case eqs. \eqref{ein} are compatible and could determine all 
unknown $a_i,\rho$ and $\lambda$. As in the previous case, the 
numerical simulations will be presented elsewhere \cite{SRVprep} and 
here we shall analyze only the asymptotic behavior of the solutions.

Let us assume that for large $t$
\begin{equation}
a_3 \sim v^\gamma \,,
\end{equation}
and consequently 
\begin{eqnarray}
a_1 &\sim& v^{\frac {n(1-2\gamma) + m\gamma}{m+n}}\,,\nonumber\\
a_2 &\sim& v^{\frac {n\gamma + m(1-2\gamma)}{m+n}}\,,
\end{eqnarray}
and it is quite simple to verify that all equations \eqref{ein} support 
this behavior in the asymptotic regime $t\longrightarrow \infty$. For a 
particular value of $\gamma$, namely for $\gamma = \frac{1}{3}$, we 
recover the isotropization \eqref{asymp} discussed above. 

Let us observe that from eq. \eqref{emc} we have in the asymptotic 
regime
\begin{equation}
\dot \rho = \frac{\dot v}{v}(-\rho + \gamma \lambda)\,.
\end{equation}

We could consider that $\rho$ has an asymptotic behavior correlated 
with that of $v$
\begin{equation}
\rho \sim v^\delta\,,
\end{equation}
with $\delta$ some constant which imply a proportionality relation 
between $\rho$ and $\lambda$ as in eq. \eqref{rhoalphalambda}, namely,
$\lambda = \frac{1+\delta}{\gamma} \rho$.

With all these assumptions a similar equation of evolution for $v$ as 
in \eqref{quad1asym} holds with appropriate constants of integration. 
The corresponding analysis of the asymptotic behavior of solutions of 
this equation proceeds as above.
\section{Conclusion}

We have studied the evolution of an anisotropic Universe given by a
BVI cosmological model in presence of a cloud of cosmic strings
and magnetic flux. It is found that the system with 
$z$-dependent magnetic field within the scope of BVI spacetime is not 
consistent in general. This difficulty could be overcome working either
in a BVI$_0$ metric, setting $m=n$, or introducing a particular
$z$-dependence of the magnetic permeability.

In a forthcoming paper \cite{SRVprep} we shall present some numerical
simulations and a detailed analysis of the stability and singularities
of the field equations for the present cosmological model.

\subsection*{Acknowledgments}

We are thankful to Suresh Kumar for helpful comments.
This work is supported in part by a joint Romanian-LIT, JINR, 
Dubna Research Project, theme no. 05-6-1060-2005/2010. M.V. is 
partially supported by CNCSIS program IDEI-571/2008 and NUCLEU 
program PN-09370102, Romania.

\setcounter{equation}{0} \renewcommand{\theequation}
{A.\arabic{equation}}
\section*{Appendix A. BVI cosmological model}

The gravitational field in the present paper is given by a
BVI cosmological model in the form \eqref{bvi}. A
suitable choice of $m,\,n$ as well as the metric functions
$a_1,\,a_2,\,a_3$ in the BVI given by \eqref{bvi} evokes the
following Bianchi-type Universes:
\begin{itemize}
\item
for $m=n$ the BVI metric transforms to a BVI$_0$ 
one, i.e., $m = n$,\\ BVI $\Longrightarrow$ BVI$_0$ $\in$ open 
FRW with the line elements
\begin{equation}
ds^2 = dt^2 - a_1^{2} e^{-2mz}\,dx^{2} - a_2^{2} e^{2mz}\,dy^{2} -
a_3^{2}\,dz^2\,; \label{bv}
\end{equation} 
\item for $m=-n$ one gets the Bianchi-type V (BV) spacetime;
\item
for $n=0$ the BVI metric transforms to a Bianchi-type III (BIII)
one, i.e., $n = 0$, BVI $\Longrightarrow$ BIII with the line
elements
\begin{equation}
ds^2 = dt^2 - a_1^{2} e^{-2mz}\,dx^{2} - a_2^{2} \,dy^{2} -
a_3^{2}\,dz^2\,; \label{biii}
\end{equation}
\item
for $m=n =0$ the BVI metric transforms to a Bianchi-type I (BI) one,
i.e., $m = n = 0$, BVI $\Longrightarrow$ BI with the line elements
\begin{equation}
ds^2 = dt^2 - a_1^{2} \,dx^{2} - a_2^{2} \,dy^{2} - a_3^{2}\,dz^2\,;
\label{bi}
\end{equation}
\item
for $m=n=0$ and equal scale factor in all three directions the BVI
metric transforms to a Friedmann-Robertson-Walker (FRW) Universe,
i.e., $m = n = 0$ and $a=b=c$, BVI $\Longrightarrow$ FRW with the
line elements
\begin{equation}
ds^2 = dt^2 - a^{2} \bigl(dx^{2} + \,dy^{2} + \,dz^2\bigr)\,.
\label{frw}
\end{equation}
\end{itemize}

Let us go back to the BVI cosmological model
\eqref{bvi}.
The nontrivial Chris\-tof\-fel symbols of the BVI metric read
\begin{eqnarray}
\Gamma_{01}^{1} &=& \frac{\dot a_1}{a_1},\quad
\Gamma_{02}^{2} = \frac{\dot a_2}{a_2},\quad
\Gamma_{03}^{3} = \frac{\dot a_3}{a_3}\,, \nonumber\\
\Gamma_{11}^{0} &=& a_1 \dot{a_1} e^{-2mz},\quad \Gamma_{22}^{0} = a_2
\dot{a_2} e^{2nz},\quad
\Gamma_{33}^{0} = a_3 \dot{a_3}\,,\nonumber\\
\Gamma_{31}^{1} &=& -m,\quad \Gamma_{32}^{2} = n,\quad \Gamma_{11}^{3}
= \frac{m a_1^2}{a_3^2} e^{-2mz},\quad \Gamma_{22}^{3}
= -\frac{n a_2^2}{a_3^2} e^{2nz}\,. \nonumber
\end{eqnarray}

The non-vanishing components of Riemann tensor corresponding to
\eqref{bvi} are
\begin{eqnarray}
R_{\,\,\,01}^{01} &=& -\frac{\ddot a_1}{a_1}, \quad
R_{\,\,\,02}^{02} = -\frac{\ddot a_2}{a_2}, \quad
R_{\,\,\,03}^{03} = -\frac{\ddot a_3}{a_3}\,, \nonumber \\
R_{\,\,\,12}^{12} &=& -\frac{mn}{a_3^2} - \frac{\dot
a_1}{a_1}\frac{\dot a_2}{a_2},\quad R_{\,\,\,13}^{13} =
\frac{m^2}{a_3^2} - \frac{\dot a_3}{a_3}\frac{\dot a_1}{a_1}, \quad
R_{\,\,\,23}^{23} = \frac{n^2}{a_3^2} - \frac{\dot
a_2}{a_2}\frac{\dot a_3}{a_3}\,, \nonumber\\
R_{\,\,\,31}^{10} &=& \frac{m}{a_3^2} \Bigl(\frac{\dot a_1}{a_1} -
\frac{\dot a_3}{a_3} \Bigr),\quad R_{\,\,\,01}^{13} = m
\Bigl(\frac{\dot a_3}{a_3} - \frac{\dot a_1}{a_1} \Bigr)\,,\nonumber\\
R_{\,\,\,32}^{20} &=& \frac{n}{a_3^2} \Bigl(\frac{\dot a_3}{a_3} -
\frac{\dot a_1}{a_1} \Bigr),\quad R_{\,\,\,02}^{23} = n
\Bigl(\frac{\dot a_2}{a_2} - \frac{\dot a_3}{a_3} \Bigr)\,. \nonumber
\end{eqnarray}

The nontrivial components of the Ricci tensor are
\begin{eqnarray}
R_{3}^{0} &=& -\Bigl(m\frac{\dot a_1}{a_1} - n\frac{\dot a_2}{a_2} -
(m -n ) \frac{\dot a_3}{a_3}\Bigr)\,, \nonumber\\
R_{0}^{0} &=& \Bigl(\frac{\ddot a_1}{a_1} + \frac{\ddot a_2}{a_2} +
\frac{\ddot a_3}{a_3}\Bigr)\,, \nonumber\\
R_{1}^{1} &=&  \Bigl(\frac{\ddot a_1}{a_1}+ \frac{\dot
a_1}{a_1}\frac{\dot a_2}{a_2} + \frac{\dot a_1}{a_1}\frac{\dot
a_3}{a_3}
-\frac{m^2 - mn}{a_3^2}\Bigr)\,, \nonumber\\
R_{2}^{2} &=&  \Bigl(\frac{\ddot a_2}{a_2}+ \frac{\dot
a_1}{a_1}\frac{\dot a_2}{a_2} + \frac{\dot a_2}{a_2}\frac{\dot
a_3}{a_3}
-\frac{n^2 - mn}{a_3^2}\Bigr)\,, \nonumber\\
R_{3}^{3} &=&  \Bigl(\frac{\ddot a_3}{a_3}+ \frac{\dot
a_1}{a_1}\frac{\dot a_3}{a_3} + \frac{\dot a_2}{a_2}\frac{\dot
a_3}{a_3} -\frac{m^2 + n^2}{a_3^2}\Bigr)\,. \nonumber
\end{eqnarray}

The Ricci scalar reads
\begin{equation}
R = 2\bigl[\frac{\ddot a_1}{a_1} + \frac{\ddot a_2}{a_2} +
\frac{\ddot a_3}{a_3} + \frac{\dot a_1}{a_1}\frac{\dot a_2}{a_2} +
\frac{\dot a_2}{a_2}\frac{\dot a_3}{a_3} + \frac{\dot
a_3}{a_3}\frac{\dot a_1}{a_1} - \frac{m^2 - mn + n^2}{a_3^2}\bigr]\,.
\end{equation}

Let us now find expansion and shear for BVI metric. The expansion is
given by
\begin{equation}
\vartheta = u^\mu_{;\mu} = u^\mu_{\mu} + \Gamma^\mu_{\mu\alpha}
u^\alpha\,, \label{expansion}
\end{equation}
and the shear is given by
\begin{equation}
\sigma^2 = \frac{1}{2} \sigma_{\mu\nu} \sigma^{\mu\nu}\,,
\label{shear}
\end{equation}
with
\begin{equation}
\sigma_{\mu\nu} = \frac{1}{2}\bigl[u_{\mu;\alpha} P^\alpha_\nu +
u_{\nu;\alpha} P^\alpha_\mu \bigr] - \frac{1}{3} \vartheta
P_{\mu\nu}\,, \label{shearcomp}
\end{equation}
where the projection vector $P$:
\begin{equation}
P^2 = P, \quad P_{\mu\nu} = g_{\mu\nu} - u_\mu u_\nu, \quad
P^\mu_\nu = \delta^\mu_\nu - u^\mu u_\nu\,. \label{proj}
\end{equation}
In comoving system we have $u^\mu = (1,0,0,0)$. In this case one
finds
\begin{equation}
\vartheta = \frac{\dot a_1}{a_1} + \frac{\dot a_2}{a_2} + \frac{\dot
a_3}{a_3}\,, \label{expbvi}
\end{equation}
and
\begin{equation}
\sigma_{11} = \frac{a_1^2 e^{-2mz}}{3} \Bigl(-2\frac{\dot a_1}{a_1}
+ \frac{\dot a_2}{a_2} + \frac{\dot a_3}{a_3}\Bigr) \Longrightarrow
\sigma^1_1 = -\frac{1}{3}\Bigl(-2\frac{\dot a_1}{a_1} + \frac{\dot
a_2}{a_2} + \frac{\dot a_3}{a_3}\Bigr) \,, \label{sh11}
\end{equation}
\begin{equation}
\sigma_{22} = \frac{a_2^2 e^{2nz} }{3} \Bigl(-2\frac{\dot a_2}{a_2}
+ \frac{\dot a_3}{a_3} + \frac{\dot a_1}{a_1}\Bigr)  \Longrightarrow
\sigma^2_2 = -\frac{1}{3}\Bigl(-2\frac{\dot a_2}{a_2} + \frac{\dot
a_3}{a_3} + \frac{\dot a_1}{a_1}\Bigr) \,, \label{sh22}
\end{equation}
\begin{equation}
\sigma_{33} = \frac{a_3^2}{3}\Bigl(-2\frac{\dot a_3}{a_3} +
\frac{\dot a_1}{a_1} + \frac{\dot a_2}{a_2}\Bigr)  \Longrightarrow
\sigma^3_3 = -\frac{1}{3}\Bigl(-2\frac{\dot a_3}{a_3} + \frac{\dot
a_1}{a_1} + \frac{\dot a_2}{a_2}\Bigr) \,. \label{sh33}
\end{equation}


%
\end{document}